\documentclass[a4paper,twocolumn,10pt,accepted=2021-07-06]{quantumarticle}
\pdfoutput=1

\usepackage[square,numbers,sort&compress]{natbib}
\usepackage[T1]{fontenc}
\usepackage[utf8]{inputenc}
\usepackage{graphicx,epsfig}
\usepackage{amsmath,amssymb,amsthm,bbm,bm}
\usepackage{import}
\usepackage{booktabs}
\usepackage{exscale}
\usepackage[english]{babel}
\usepackage{color}
\usepackage{appendix}
\usepackage{dsfont}
\usepackage{hyperref}
\usepackage{xcolor}
\hypersetup{
	colorlinks,
	linkcolor={rgb:red,2; blue,1},
	citecolor={blue!60!black},
	urlcolor={blue!60!black}
}

\renewcommand{\H}{\hat{H}}

\newcommand{\ii}{\ensuremath{\mathrm{i}}}
\newcommand{\ee}{\ensuremath{\mathrm{e}}}
\newcommand{\dd}{\mathrm{d}}

\newcommand{\od}[2]{\ensuremath{\frac{\dd #1}{\dd #2}}}

\newcommand{\bra}[1]{\ensuremath{\left\langle #1 \right\rvert}}
\newcommand{\ket}[1]{\ensuremath{\left\lvert #1 \right\rangle}}

\newcommand{\ketbra}[2]{\ensuremath{\left| #1 \middle\rangle\middle\langle #2\right|} }

\newcommand{\avg}[1]{\left \langle #1 \right \rangle}

\newcommand{\Tr}{\mathrm{Tr}}

\newcommand{\sg}{\ensuremath{\hat{\sigma}}}

\renewcommand{\H}{\ensuremath{\hat{H}}}
\newcommand{\DD}{\ensuremath{\mathcal{D}}}
\newcommand{\HH}{\ensuremath{\mathcal{H}}}

\graphicspath{{./figs/}}

\begin{document}
\title{Charging a quantum battery with linear feedback control}
\author{Mark T. Mitchison}
\email{mark.mitchison@tcd.ie}
\affiliation{School of Physics, Trinity College Dublin, College Green, Dublin 2, Ireland}
\author{John Goold}
\affiliation{School of Physics, Trinity College Dublin, College Green, Dublin 2, Ireland}
\author{Javier Prior}
\affiliation{Departamento de F\'isica Aplicada, Universidad Polit\'ecnica de Cartagena, Cartagena E-30202, Spain}
\affiliation{Instituto Carlos I de F\'isica Te\'orica y Computacional, Universidad de Granada, Granada E-18071, Spain}

\begin{abstract}
Energy storage is a basic physical process with many applications. When considering this task at the quantum scale, it becomes important to optimise the non-equilibrium dynamics of energy transfer to the storage device or battery. Here, we tackle this problem using the methods of quantum feedback control. Specifically, we study the deposition of energy into a quantum battery via an auxiliary charger. The latter is a driven-dissipative two-level system subjected to a homodyne measurement whose output signal is fed back linearly into the driving field amplitude. We explore two different control strategies, aiming to stabilise either populations or quantum coherences in the state of the charger. In both cases, linear feedback is shown to counteract the randomising influence of environmental noise and allow for stable and effective battery charging. We analyse the effect of realistic control imprecisions, demonstrating that this good performance survives inefficient measurements and small feedback delays. Our results highlight the potential of continuous feedback for the control of energetic quantities in the quantum regime. 
\end{abstract}
\maketitle

\section{Introduction}

The ability to control the quantum dynamics of mesoscopic systems has opened a broad research frontier in the physical sciences, spanning metrology~\cite{Wiseman2009}, information processing~\cite{Nielsen2010}, and non-equilibrium statistical mechanics~\cite{Polkovnikov2011}. With the growing complexity of experiments in these fields, a thorough understanding of the energetics of quantum systems is increasingly important, both as a diagnostic tool and to help design optimal control protocols. A fruitful way to gain insight is through specific examples of thermodynamic processes in the quantum regime, such as energy storage and extraction. To that end, quantum batteries  --- dynamical systems that receive and supply energy --- have emerged as a useful paradigm to explore the fundamental limits and potential benefits of energy transduction with quantum degrees of freedom~\cite{Campaioli2018,Bhattacharjee2020}.

Even far from thermal equilibrium, the second law of thermodynamics (or an appropriately generalised version thereof) constrains the amount of useful work that can be extracted from a quantum system~\cite{Allahverdyan2004,Esposito2011,Brandao2015,Brown2016,Landi2020}. Nevertheless, various strategies have been developed to boost the extractable work, such as operating collectively on several systems~\cite{Alicki2013,Hovhannisyan2013} and exploiting quantum correlations~\cite{PerarnauLlobet2015,Giorgi2015} or coherences~\cite{Korzekwa2016,Francica2020}. Likewise, it has been shown that collective operations can enhance the charging power of composite quantum batteries~\cite{Binder2015,Campaioli2017}. These predictions have inspired a substantial body of theoretical research aiming to harness quantum or many-body effects in order to improve the performance of energy storage devices. Numerous quantum battery architectures have since been proposed~\cite{Le2018,Andolina2019,Andolina2019a,Ferraro2019,Rosa2020,Zhang2019,Caravelli2020,Ghosh2020,Rossini2020} and the effects of different physical phenomena --- ranging from entanglement~\cite{Kamin2020,JuliaFarre2020} to many-body localisation~\cite{Rossini2019} --- have been extensively investigated.

One phenomenon that is especially relevant for practical energy storage is dissipation stemming from interactions with the environment. A good battery should be well isolated from its surroundings in order to prevent the loss of charge over time~\cite{Pirmoradian2019,Kamin2020a,Carrega2020}. Yet a perfectly isolated battery is decoupled from all external energy sources and thus cannot be charged in the first place. A natural way to resolve this dilemma is to ensure that the power supply is physically separated from the system used for long-term energy storage. This can be achieved by supplying energy to an auxiliary system --- the charger --- which is allowed to interact with the battery in a controlled way~\cite{Andolina2018, Ferraro2018}. However, the coupling between the charger and its external power supply necessarily introduces noise, which limits the charging process~\cite{Farina2019}.  Recent proposals have shown how to mitigate the effect of environmental noise via measurements~\cite{Gherardini2020} or dark states~\cite{Santos2019a, Quach2020}, while other authors have instead suggested to harness noise as a charging mechanism~\cite{Barra2019,Latune2019,Hovhannisyan2020,Cakmak2020,Tabesh2020,Tacchino2020}. These approaches are notable due to their stability, meaning that the battery's charge tends to a stationary value instead of oscillating over time~\cite{Santos2019a,Santos2020}. This desirable feature mimics the behaviour of battery charging in everyday life and removes the need for precisely timed switching of the battery-charger coupling.

Here, we propose an alternative route based on continuous weak measurements and feedback control. This effectively recasts the problem of dissipative battery charging as a feedback stabilisation protocol~\cite{Hofmann1998,Wang2001,Ruskov2002,Patti2017}. More precisely, we consider a homodyne-like measurement scheme, which leads to a dynamical description in terms of diffusive quantum trajectories~\cite{Belavkin1987, Wiseman1993,*Wiseman1994,Korotkov1999}. This framework applies to a variety of experimental settings --- including optical, atomic, and electronic systems --- where continuous quantum feedback control has already been implemented~\cite{Serafini2012,Zhang2017}. The thermodynamics of such feedback has recently been studied both theoretically~\cite{Horowitz2015,Alonso2016,Elouard2017} and experimentally~\cite{Naghiloo2020,Debiossac2020}.

In order to explore the effect of feedback on the charging of an open quantum battery, we adopt a simple model that is analytically and numerically tractable, as detailed in Sec.~\ref{sec:model}. We consider a two-level charger coupled to a finite-dimensional battery and specialise to direct, linear feedback, where the driving strength is directly proportional to the measurement signal. In Sec.~\ref{sec:results}, we introduce and analyse two different control protocols, based on stabilising either populations or coherences in the charger. Despite its simplicity, linear feedback is shown to enable highly stable and effective energy transfer from the charger to the battery. In the ideal case of efficient measurements and instantaneous feedback, the battery can be charged perfectly (i.e.~to its maximum-energy state). We also analyse the effect of measurement inefficiency and feedback delay in detail, finding good performance even in the presence of realistic imperfections. We discuss our results and suggest interesting future directions in Sec.~\ref{sec:conclusion}. The Appendix provides an analysis of the effect of thermal noise and detuning between the charger and battery. Units where $\hbar=1$ are used throughout.

\section{Model}
\label{sec:model}
\subsection{Description of the system}

The objective is to pump energy into a $d$-level quantum battery, $B$, via a two-level system (qubit), $C$, which acts as the charger. The battery is modelled as a ladder of equidistant states separated by energy $\omega_0$, with the same energy splitting characterising the charger. The bare Hamiltonian is thus given by $\H_0 = \H_C + \H_B$, with
\begin{equation}
\label{H0}
\H_C = \frac{\omega_0}{2} \sg_z, \qquad \H_B = \omega_0\hat{N},
\end{equation}
where $\sg_{x,y,z}$ are standard Pauli operators describing the qubit and we defined the number operator of the battery
\begin{align}
\label{BnumberOp}
\hat{N} &= \sum_{n=0}^{d-1} n\ketbra{n}{n}.
\end{align}
During the charging process, the charger resonantly exchanges energy with the battery via the interaction Hamiltonian
\begin{equation}\label{qubit_battery_interaction}
\hat{H}_{\rm int} = g\left (\sg_+ \hat{B} + \sg_- \hat{B}^\dagger \right ),
\end{equation}
where $\sg_\pm = \tfrac{1}{2}(\sg_x\pm \ii \sg_y)$ are raising and lowering operators for the qubit, $g$ is the coupling strength, and we defined the lowering operator for the battery
\begin{equation}\label{battery_ladder}
\hat{B} = \sum_{n=1}^{d-1}\ket{n-1}\bra{n}.
\end{equation}
Note that, since $[\H_0, \H_{\rm int}] = 0$, the interaction can be switched on and off without affecting the average local energy of the charger and battery, $\langle \hat{H}_0\rangle$, in principle~\cite{Andolina2018}.

\subsection{Open-system dynamics and feedback control}
\label{sec:feedback_dynamics}

The battery is assumed to be a well isolated system, so that its direct interaction with the environment can be neglected on the timescales under consideration. In contrast, the charger is an open quantum system that couples to an external field. This coupling allows energy to be pumped into the charger by a coherent driving tone, but also necessarily entails dissipation due to the many-mode field acting as a reservoir. We focus on the low-temperature regime, where the thermal energy is much less than $\omega_0$ so that the probability of photon absorption from the reservoir field is negligible. Working in an interaction picture with respect to $\H_0$ and invoking the rotating-wave approximation, the coherent drive is described by the Hamiltonian
\begin{equation}
	\label{driving_Hamiltonian}
	\H_{\rm drive}(t) = \Omega(t)\sg_y,
\end{equation}
where the Rabi frequency $\Omega(t)$ is proportional to the intensity of the driving field. In the absence of any feedback, the driven-dissipative dynamics is then described by the master equation 
\begin{equation}
	\label{ME_no_measurement}
	\frac{\dd \hat{\rho}}{\dd t} = -\ii [\H(t), \hat{\rho}(t)] + \Gamma \mathcal{D}[\sg_-] \hat{\rho}(t),
\end{equation}
where $\H(t) = \H_{\rm int} + \H_{\rm drive}(t)$, $\Gamma$ is the spontaneous emission rate, and $\DD[\hat{A}]\bullet = \hat{A}\bullet\hat{A}^\dagger - \tfrac{1}{2}\{\hat{A}^\dagger\hat{A},\bullet\}$ defines a Lindblad dissipator. Eq.~\eqref{ME_no_measurement} is based on the Born-Markov and rotating-wave approximations, where the latter requires $\omega_0$ to be the largest energy scale in the system. In particular, we assume that $\omega_0 \gg |\Omega|, g,\Gamma$, and that $\Omega(t)$ varies slowly compared to the fast timescale $\omega_0^{-1}$~\footnote{This assumption is consistent with the feedback protocol introduced in Eq.~\eqref{feedback_drive} so long as the white noise in Eq.~\eqref{measRec} is understood as an idealisation of the true noise at the detector output, which of course has a finite correlation time $t_c$. The rotating-wave approximation is justified so long as $\omega_0 \gg t_c^{-1} \gg \Gamma, g, |f|,|\Omega_0|$}.

An optimal driving protocol $\Omega(t)$ should maximise the energy deposited in the battery that can subsequently be extracted. A simple approach is to drive the qubit with a constant intensity, i.e.~$\Omega(t) = \Omega_0 = \rm const.$, as considered in Ref.~\cite{Farina2019}. However, this strategy is only effective at charging the battery transiently, with the extractable work oscillating over time and eventually decaying to a fraction of the maximum. We note that similar transient oscillatory effects have been identified in small absorption refrigerators~\cite{Mitchison2015,Brask2015,Nimmrichter2017,Maslennikov2019}. The ineffectiveness of the charger at long times is caused by spontaneous emission, which randomises the phase coherence of the driven qubit and prevents a stationary population inversion from being established. This motivates our alternative approach, where phase information that is lost to the environment is partially regained by a weak measurement and then fed back into the control field. 

\begin{figure}
	\includegraphics[width=\linewidth]{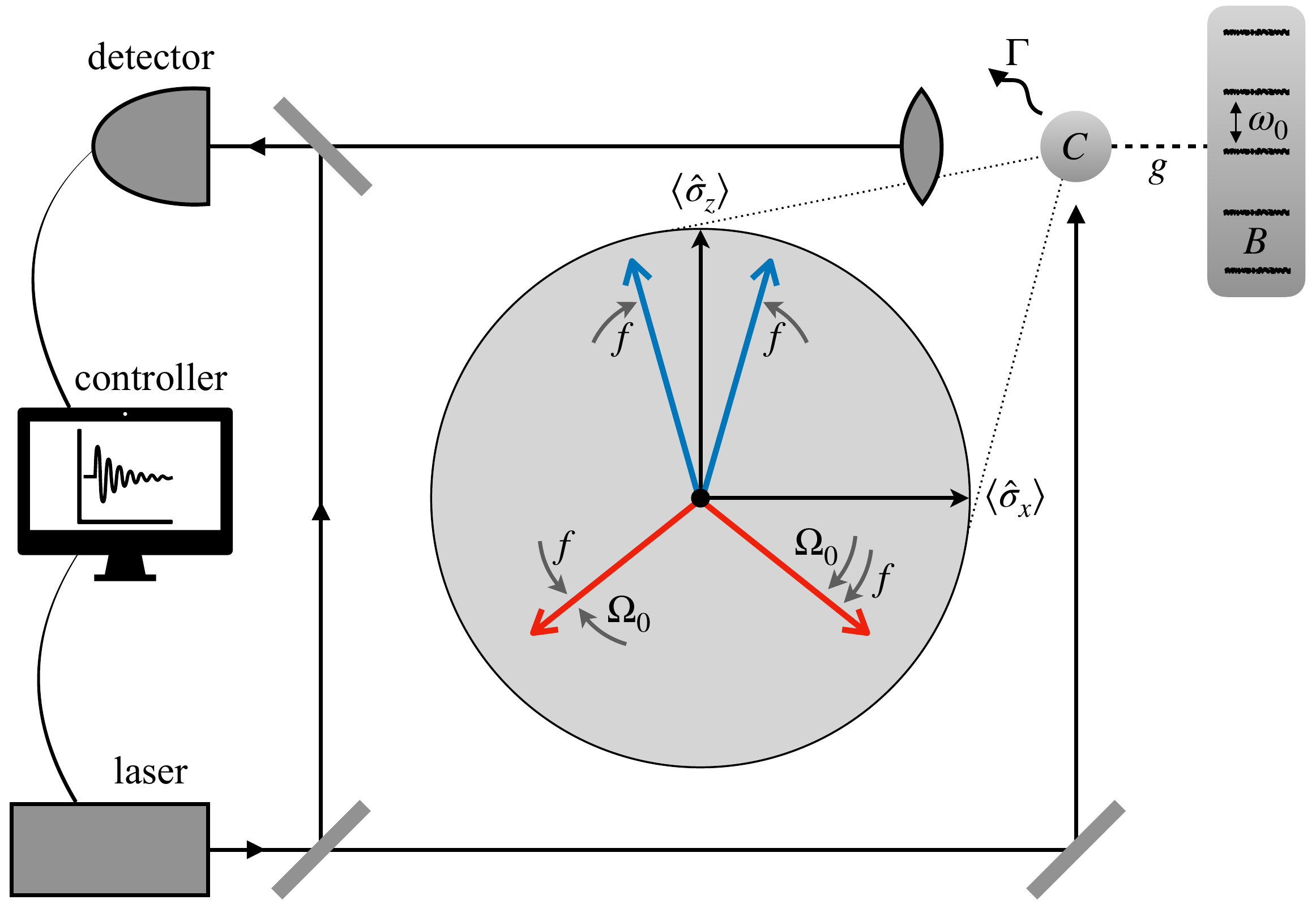}
	\caption{Illustration of the feedback loop in a quantum-optics setting. Laser light drives a two-level system, $C$, coupled to a $d$-level battery, $B$. Spontaneously emitted photons are collected by a lens and interfered with light from the laser source in a homodyne measurement. The measurement record is used by the controller to modulate the laser intensity. The central inset depicts two different feedback control strategies on the Bloch sphere ($x$--$z$ plane) of the qubit. Stabilising the Bloch vector near the vertical (blue arrows) charges the battery by inducing population inversion, while stabilising in the left hemisphere (red arrows) generates coherence.
		\label{fig:schematic}}
\end{figure}

We assume that some of the spontaneously emitted photons are collected and measured with a homodyne interferometer, as shown in Fig.~\ref{fig:schematic}. The total measurement efficiency is denoted by $\eta\leq 1$, which incorporates both the fraction of collected photons, $\eta_{\rm c}$, and the detector efficiency, $\eta_{\rm d}$, so that $\eta = \eta_{\rm c} \eta_{\rm d}$. The resulting homodyne current is represented by an appropriately normalised and shifted measurement record~\cite{Carmichael1993,Wiseman1993,Wiseman1994}
\begin{equation}
	\label{measRec}
	r(t)\dd t = \Tr\{\hat{\rho}_r(t) \sg_x\} \dd t + \frac{\dd w(t)}{\sqrt{\eta \Gamma}},
\end{equation}
where $\hat{\rho}_r(t)$ is the quantum state (in the interaction picture) conditioned on the measurement record and $\dd w(t)$ is a Wiener increment that represents noise in the detector output. The average measurement signal yields an estimate of the qubit coherence as $\mathbb{E}[r(t)] = \avg{\sg_x}$, where $\mathbb{E}[\bullet]$ denotes an average over the noise~\footnote{Note that we assume that the measurement apparatus is arranged so that the measured field quadrature is orthogonal to the one driving the system. This definite phase relationship requires both fields to derive from the same source, as indicated in Fig.~\ref{fig:schematic}.}. We emphasise that the optical apparatus depicted in Fig.~\ref{fig:schematic}, similar to optomechanical realisations~\cite{Bushev2006,Tebbenjohanns2019,Conangla2019}, is selected mainly for illustrative purposes. Analogous continuous weak measurements have been implemented on other platforms, for example, using microwave fields~\cite{Vijay2012,Naghiloo2020} or electrical currents~\cite{Buks1998,Korotkov1999, Goan2001}. 

Feedback is enacted by applying a driving field that depends on the results of the measurement. We consider the simplest case of \textit{direct} feedback, in which the drive intensity is proportional to the measurement record, i.e.
\begin{equation}
	\label{feedback_drive}
	\Omega(t) = \Omega_0 - f r(t-\tau),
\end{equation}
where $\Omega_0$ is a constant drive, $f$ parametrises the strength of feedback, and we have allowed for a small time delay $\tau>0$ in the feedback loop. Since feedback is applied after the measurement record is read out, the evolution is described by the map~\cite{Wiseman2009}
\begin{equation}\label{drive_after_measure}
	\hat{\rho}_r(t+\dd t) = \ee^{\HH[-\ii \hat{H}(t)]\dd t} \big (\hat{\rho}_r(t) + \mathcal{K} [\hat{\rho}_r(t) ]\big ),
\end{equation}
where $\mathcal{K}[\hat{\rho}_r(t)]$ represents the dissipative part of the evolution
\begin{align}\label{heat_increment}
	\mathcal{K} [\hat{\rho}_r(t) ]  & = \Gamma \DD[\sg_-] \hat{\rho}_r(t)\dd t + \sqrt{\eta\Gamma} \HH[\sg_-]\hat{\rho}_r(t)\dd w(t), 
\end{align}
while $\HH[\hat{A}] \bullet = \hat{A} \bullet + \bullet \hat{A}^\dagger -\Tr [\bullet (\hat{A} + \hat{A}^\dagger)]\bullet$ is the innovation superoperator and $\dd w(t)$ is the same Wiener increment appearing in Eq.~\eqref{measRec}. 

Expectation values are computed from the ensemble-averaged density matrix $\hat{\rho}(t) = \mathbb{E}[\hat{\rho}_r(t)]$. In the ideal case of negligible feedback delay, $\tau\to 0$, it is possible to recover a Markovian master equation for $\hat{\rho}(t)$. In particular, by expanding Eq.~\eqref{drive_after_measure} to first order in $\dd t$ while applying the rules of It\^{o} calculus (i.e.~$\dd w^2 = \dd t$), one obtains~\cite{Wiseman2009}
\begin{align}\label{unconditional_master_equation}
	\od{\hat{\rho}}{t} & = -\ii [\H_{\rm int} + \Omega_0\sg_y,\hat{\rho}] + \ii f [\sg_y, \sg_- \hat{\rho} + \hat{\rho} \sg_+] \notag \\
	&\quad  + \Gamma \DD[\sg_-] \hat{\rho}  + \frac{f^2}{\eta\Gamma} \DD[\sg_y]\hat{\rho}.
\end{align}
The first line above describes coherent evolution due to interactions and driving, while the second line describes the effect of spontaneous emission and the measurement noise that  the drive feeds back into the system. For finite values of $\tau$, however, no such Markovian description is possible and it is necessary to solve the explicit It\^{o} equation~\eqref{drive_after_measure} and average over many trajectories. 

It is straightforward to extend our model to describe finite-temperature dissipation and detuning between the charger and the battery. These effects are analysed in detail in the Appendix.

\subsection{Energetics of the battery}

At the end of the charging process, the charger is decoupled from the battery. The main energetic quantity of interest is then how much energy can subsequently be extracted from the battery as work. The mean energy of the battery is denoted by $E[\hat{\rho}_B] = \Tr[\hat{\rho}_B \H_B]$, where $\hat{\rho}_B = \Tr_C[\hat{\rho}]$ is the battery's state, which may be far from equilibrium and exhibit significant energy fluctuations and entropy. As a result, the extractable work is generally less than $E[\hat{\rho}_B]$.

In order to quantify the useful portion of the deposited energy, we use the ergotropy~\cite{Allahverdyan2004}, which upper-bounds the work that can be extracted from the battery by a cyclic variation of its Hamiltonian. Any such cyclic process generates a unitary operation $\hat{U}$, which reduces the average energy of the battery by an amount
\begin{equation}\label{work_extracted}
	W_{\rm ex} = E [\hat{\rho}_B ] - E [\hat{U} \hat{\rho}_B \hat{U}^\dagger].
\end{equation}
Since the unitary transformation is isentropic, an energy change $W_{\rm ex}>0$ may be attributed entirely to extracted work. The ergotropy is defined as the maximum work extractable from the state $\hat{\rho}_B$ by any unitary, i.e.~$\mathcal{E} = \max_{\hat{U}} W_{\rm ex}$. 

To get an explicit form for the ergotropy, we write the state in its eigenbasis $\{\ket{\psi_n}\}$, as $\hat{\rho}_B = \sum_n p_n \ket{\psi_n}\bra{\psi_n}$, with eigenvalues ordered so that $p_{n+1} \leq p_n$. The maximal work is extracted when $\hat{U}$ takes $\hat{\rho}_B$ to the corrresponding passive state~\cite{Pusz1978}, $\hat{\pi}_{\hat{\rho}_B}  = \sum_n p_n \ket{n}\bra{n}$, where $\ket{n}$ are the eigenvectors of $\H_B$ ordered by increasing energy, i.e. $\hat{N}\ket{n} = n\ket{n}$. The ergotropy is thus given by 
\begin{equation}\label{ergotropy_explicit}
	\mathcal{E} [\hat{\rho}_B ] = E [\hat{\rho}_B ] - E [\hat{\pi}_{\hat{\rho}_B}  ].
\end{equation}

By definition, a passive state is diagonal in the energy eigenbasis, with more population in low-energy eigenstates than high-energy ones. Thermal equilibrium states are passive, for example. According to Eq.~\eqref{ergotropy_explicit}, any state possessing ergotropy must be non-passive, i.e.~having population inversion or coherence in the energy eigenbasis. Ergotropy thus quantifies the degree to which the charger-mediated energy transfer is ordered (work-like), rather than entropic (heat-like)~\cite{Binder2015a}.

Following Ref.~\cite{Francica2020}, we may explicitly distinguish the incoherent and coherent contributions to the ergotropy, so that $\mathcal{E}  = \mathcal{E}_i + \mathcal{E}_c$. The incoherent ergotropy $\mathcal{E}_i$ can be defined operationally as the maximum work extractable by a coherence-preserving unitary operation, and it is specified by $\mathcal{E}_i[\hat{\rho}_B] = \mathcal{E}[\hat{\delta}_{\hat{\rho}_B}]$, with $\hat{\delta}_{\hat{\rho}_B} = \sum_n \ketbra{n}{n} \hat{\rho}_B \ketbra{n}{n}$ the dephased state in the energy eigenbasis. Because $\mathcal{E}_i$ depends only on the energy distribution of the state $\hat{\rho}_B$, it quantifies the work that is extractable solely by changing the populations in the energy eigenbasis. The remainder $\mathcal{E}_c = \mathcal{E} - \mathcal{E}_i$ thus isolates the contribution to ergotropy from coherence. 

\section{Results}
\label{sec:results}

\subsection{Optimal Markovian feedback}
\label{sec:optimal_control}

We begin by considering the ideal scenario of feedback with negligible delay, $\tau\to 0$. The ensemble dynamics in this case is defined by the Markovian master equation~\eqref{unconditional_master_equation}. Our aim is to choose the control parameters $\Omega_0$ and $f$ in order to maximise the final battery charge. In the following, we introduce two strategies --- appropriate for different parameter regimes --- based either on stabilising population inversion or coherence in the qubit charger.

\subsubsection{Stabilising population inversion}

\label{sec:pop_inversions}

The first control strategy aims to stabilise the state of the qubit charger as close to its excited state as possible~\cite{Wang2001}. This aim is met by setting $\Omega_0 = 0$ and $f>0$ in Eq.~\eqref{feedback_drive}, so that the Rabi frequency tends to have the opposite sign to the measurement record~\eqref{measRec}. The feedback mechanism can be understood intuitively by visualising the state of the qubit on the Bloch sphere, as depicted in the central inset of Fig.~\ref{fig:schematic} (blue arrows). Whenever the Bloch vector rotates away from the vertical, the conditional state acquires a finite expectation value $\avg{\sg_x}_r = \Tr[\sg_x\hat{\rho}_r(t)]$, which is recorded in the homodyne signal. In response, the feedback applies a drive proportional to $-f\avg{\sg_x}_r$ that acts as a restoring force.

To find the optimal value of $f$, we consider the state of the system at asymptotically long times, which is given by the stationary solution of the master equation~\eqref{unconditional_master_equation}, i.e.~$\dd \hat{\rho}/\dd t=0$. Following Ref.~\cite{Brunner2012}, we posit a product ansatz $\hat{\rho} = \hat{\rho}_C \hat{\rho}_B$ for the stationary state, where $\hat{\rho}_C$ and $\hat{\rho}_B$ are diagonal in the energy eigenbasis, viz.
\begin{align}
	\label{rho_C_stationary}
	\hat{\rho}_C  = \frac{1}{2}\left(1+\avg{\sg_z}\sg_z\right), \quad 
	\hat{\rho}_B =  \sum_{n=0}^{d-1} p_n \ket{n}\bra{n},
\end{align}
with successive populations related by a fixed ratio
\begin{equation}\label{pop_ratio}
	R = \frac{p_{n+1}}{p_n} = \frac{1+\avg{\sg_z}}{1-\avg{\sg_z}}.
\end{equation}
It is easy to check that such a state commutes with the interaction Hamiltonian, $[\H_{\rm int},\hat{\rho}_C\hat{\rho}_B] = 0$. Writing Eq.~\eqref{unconditional_master_equation} as $\dd \hat{\rho}/\dd t = -\ii [\H_{\rm int},\hat{\rho}] + \mathcal{L}_C\hat{\rho}$, the stationary state is then obtained by choosing $\hat{\rho}_C$ so that $\mathcal{L}_C\hat{\rho}_C=0$, with $\mathcal{L}_C$ acting only on the qubit degrees of freedom. The solution of this equation for $\Omega_0=0$ is fully characterised by the expectation value
\begin{equation}\label{sigma_z_stationary}
	\avg{\sg_z} = \frac{2f-\Gamma}{\Gamma - 2f +2f^2/\eta\Gamma},
\end{equation}
which is maximised by the choice $f=\Gamma$. This condition is quite intuitive, as it balances the rate of dissipation, which tends to destroy population inversion, and the restorative effect of feedback. Setting $f=\Gamma$ into Eq.~\eqref{sigma_z_stationary} yields the maximum population inversion achievable through feedback:
\begin{equation}\label{sigma_max}
	\avg{\sg_z}_{\rm max} = \frac{\eta }{2-\eta}.
\end{equation}
The maximum asymptotic charge of the battery can now be found directly from Eqs.~\eqref{rho_C_stationary}, \eqref{pop_ratio}, and \eqref{sigma_max}. The corresponding energy and ergotropy are given by
\begin{align}
	\label{energy_stationary}
	E &= \omega_0 \left [ \frac{d R^d}{R^d-1} - \frac{R	}{R-1}\right ] \\
	\label{ergotropy_stationary}
	\mathcal{E} & = \begin{cases}
		2 E - \omega_0(d-1) & (R>1) \\
		0 & (R \leq 1),
	\end{cases}
\end{align}
with $R = (1-\eta)^{-1}$ for the optimal feedback, $f=\Gamma$. 

The above solution can be interpreted in terms of thermalisation between the charger and the battery at a virtual temperature given by $T_v = -\omega_0/\ln R$~\cite{Brunner2012,Skrzypczyk2015}. For any non-zero efficiency we have $R>1$ and $T_v<0$, implying that the battery is placed in a population-inverted state with finite ergotropy. This occurs irrespectively of the value of $g$, as a consequence of the property $[\H_{\rm int},\H_{0}] = 0$ of the interaction Hamiltonian~\eqref{qubit_battery_interaction}. The maximum possible ergotropy is $E_{\rm max} = \omega_0(d-1)$, which we adopt as a convenient reference energy scale. For concreteness, we take a particular representative value $d=20$ in most of the following examples. Choosing another value of $d$ would merely rescale the final battery charge and charging time; see the Appendix for a detailed analysis.

\begin{figure}
	\centering
	\includegraphics[width=0.9\linewidth]{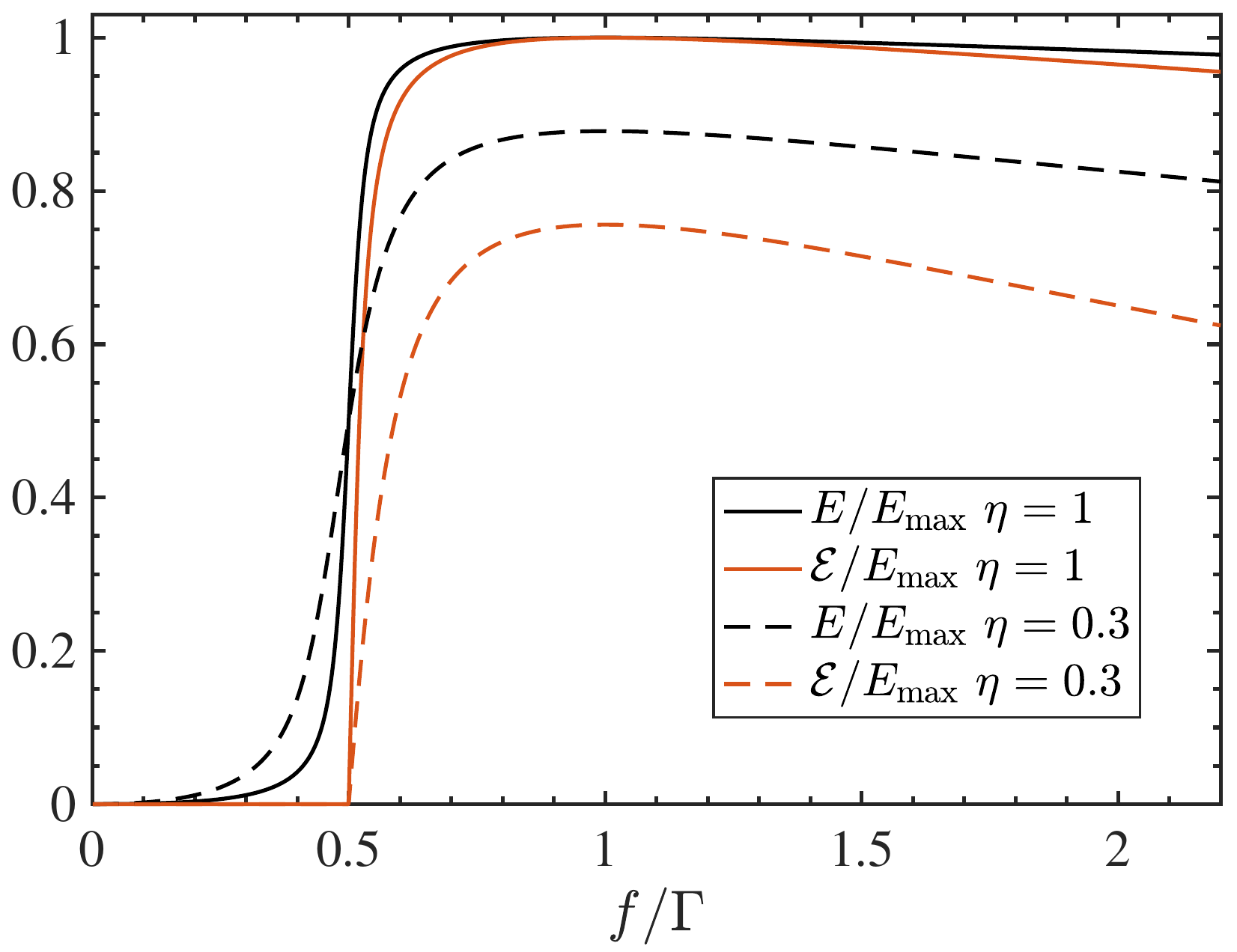}
	\caption{Asymptotic energy (black) and ergotropy (red) of the battery as a function of $f/\Gamma$, for $\Omega_0 = 0$. The maximum is always obtained at the optimal control point $f=\Gamma$. For efficient measurements (solid lines), the energy and ergotropy coincide at their maximal value, indicating perfect charging to the battery's highest excited state. For inefficient measurements (dashed lines), the difference between energy and ergotropy indicates that a portion of the deposited energy cannot be unitarily extracted as work.
		\label{fig:asymptotic_charge}}
\end{figure}

In Fig.~\ref{fig:asymptotic_charge} we plot the steady-state battery energy and ergotropy as a function of the feedback strength. We see that efficient measurements allow for perfect charging, in the sense that $\mathcal{E} = E = E_{\rm max}$ at the maximum where $f=\Gamma$. Crucially, however, Fig.~\ref{fig:asymptotic_charge} shows that rather inefficient measurements with $\eta=0.3$ still lead to significant energy and ergotropy deposited in the battery. Therefore, even when the majority of spontaneously emitted photons are irreversibly lost to the environment, it remains possible to exploit the information gained from the weak measurement to stabilise the battery in a charged state. Since this state is diagonal in the battery's Hamiltonian eigenbasis, the ergotropy $\mathcal{E} = \mathcal{E}_i$ is purely incoherent in this case.

Another notable feature of Fig.~\ref{fig:asymptotic_charge} is the sharp disappearance of the ergotropy at $f=\Gamma/2$. This reflects the fact that $\avg{\sg_z}$ changes sign to become negative for $f<\Gamma/2$, as can be seen from Eq.~\eqref{sigma_z_stationary}. Therefore, the population inversion of the charger and, correspondingly, the ergotropy of the battery both vanish when the drive is too weak. In the following section, we consider an alternative approach that is appropriate for this weak-driving regime. 

\subsubsection{Stabilising coherence}
\label{sec:stabilising_coherence}

The second control strategy targets the coherence of the qubit charger. We will see that this approach allows for charging even when the driving Rabi frequency is much smaller than the dissipation rate. We therefore restrict our considerations to the regime where $|f|<\Gamma/2$, in which we find numerically that population inversion cannot be generated. Nevertheless, by choosing $\Omega_0\neq 0$ and $f<0$, it is possible to stabilise qubit states with finite coherence in the lower half of the Bloch sphere~\cite{Hofmann1998,Wang2001}. An intuitive picture of this feedback mechanism can be understood by inspecting the inset of Fig.~\ref{fig:schematic} (red arrows). The constant drive and the conditional feedback either counterbalance or reinforce each other depending on whether the Bloch vector lies in the left or right hemisphere, generating a finite value of $\avg{\sg_x}$ on average.

\begin{figure}
	\includegraphics[width=0.9\linewidth]{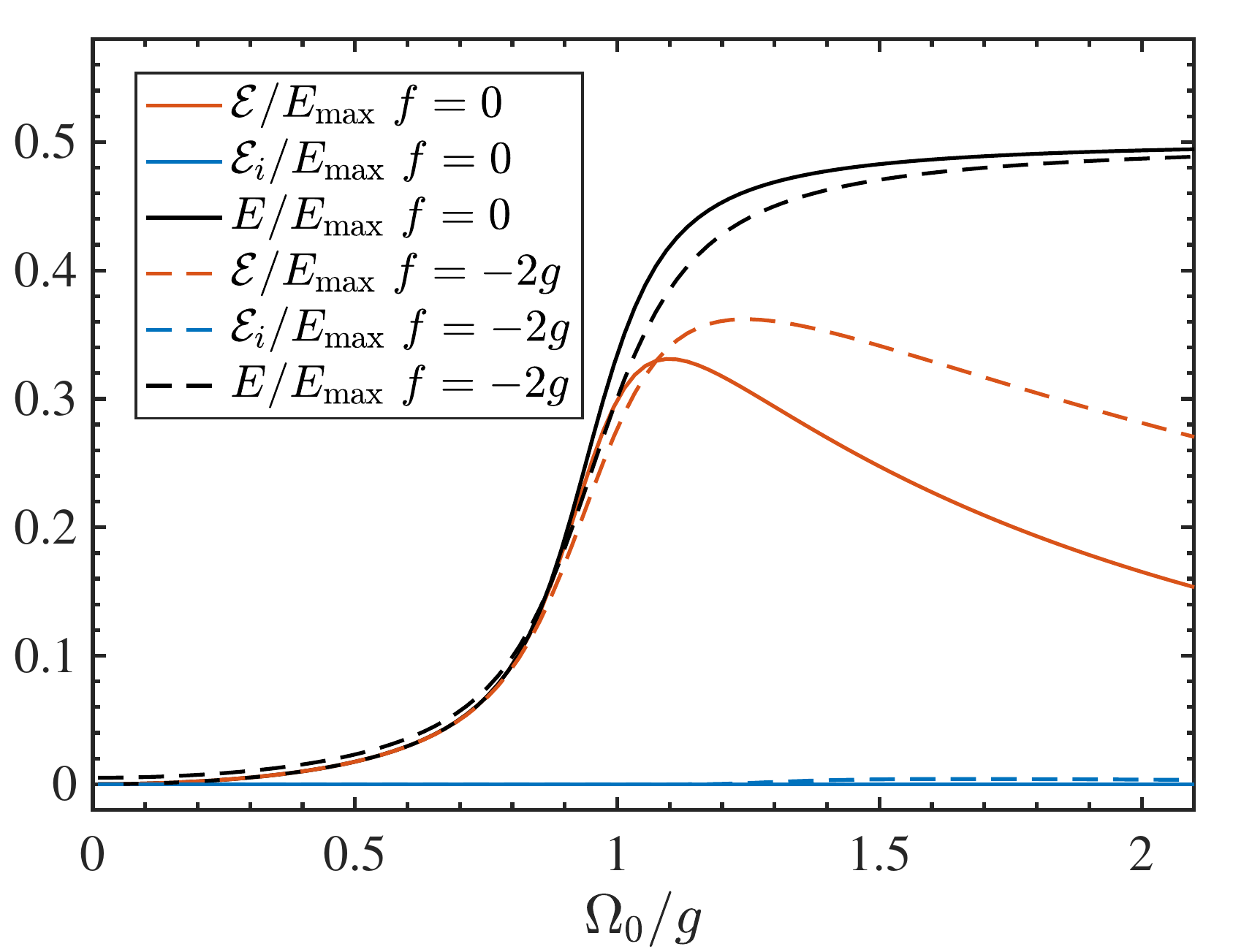}
	\caption{Asymptotic energy (black) and ergotropy (red) obtained by numerical solution of the master equation~\eqref{unconditional_master_equation} for $\Gamma = 10g$, $\eta=0.3$ and $d=20$, as a function of the unconditional Rabi frequency $\Omega_0$. Including feedback with $f<0$ leads to an increase in the maximum ergotropy (dashed lines). The incoherent ergotropy (blue) is negligibly small, indicating that the extractable work is stored almost purely in quantum coherence.
		\label{fig:final_charge_coherent}} 
\end{figure}

Unlike in the previous section, an analytical derivation of the optimal feedback parameters for arbitrary $d$ is hindered by the presence of coherences and correlations between the charger and the battery. We therefore proceed numerically by finding the zero eigenvector of the generator in Eq.~\eqref{unconditional_master_equation}, corresponding to the stationary solution of the master equation. In Fig.~\ref{fig:final_charge_coherent}, we plot the asymptotic energy and ergotropy as a function of $\Omega_0$ for an example with $\Gamma = 10g$ and $\eta = 0.3$. We observe that the optimal drive strength is on the order of the coupling $g$. Crucially, adding feedback with $f<0$ increases the maximum ergotropy that can be stored in the battery, even for inefficient measurements. We have found that larger values of $|f|$ lead to an increase in the peak ergotropy, and similar behaviour is found for other parameter choices satisfying $\Gamma \gg g,f,\Omega_0$. However, the attainable ergotropy is generally small in comparison to the control strategy discussed in Sec.~\ref{sec:pop_inversions}. 
 
Perhaps unsurprisingly, stabilising the charger's coherence leads to a build-up of almost purely coherent ergotropy in the battery. This can be seen from the blue lines in Fig.~\ref{fig:final_charge_coherent}, which demonstrate that the incoherent ergotropy $\mathcal{E}_i$ is negligibly small or zero and thus $\mathcal{E}\approx \mathcal{E}_c$. Interestingly, the inclusion of feedback leads to a small decrease in the battery's mean energy, even though the ergotropy is increased. This shows that the primary advantage of feedback in this case is to increase the purity and coherence of the battery's state. 

\subsection{Dynamics of the charging process}
\label{sec:charging_power}

\begin{figure}
	\centering
	\includegraphics[ width=0.9\linewidth]{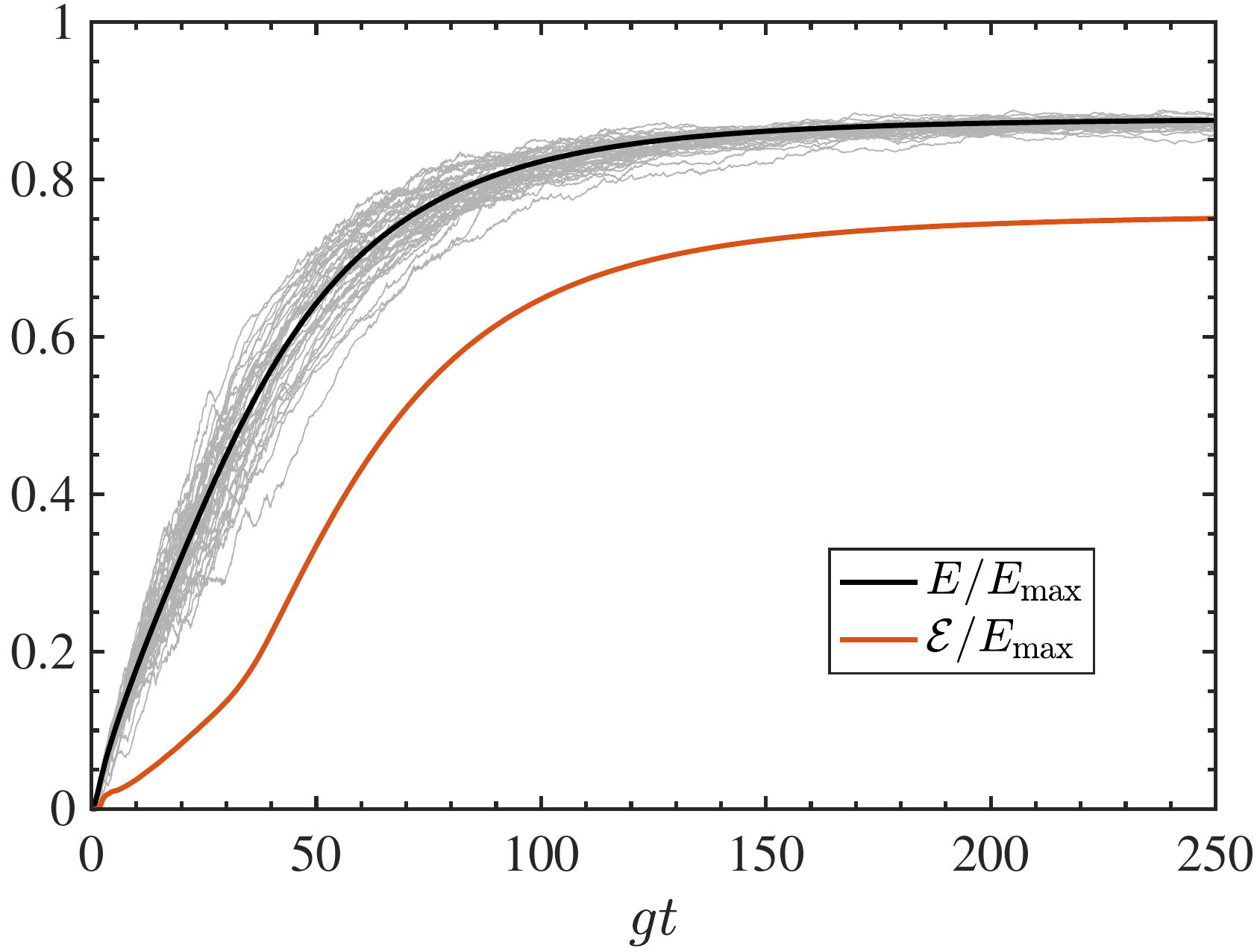}
	\caption{Dynamics of battery charging under optimal feedback stabilisation of the qubit's population inversion, with $f=\Gamma=g$, $\eta=0.3$, and $d=20$. The battery's energy is plotted against time for forty random trajectories (light grey lines) together with the energy (black line) and ergotropy (red line) of the ensemble-averaged state. 
		\label{fig:charge_dynamics}}
\end{figure}

In the previous section we discussed optimal feedback strategies that maximise the final battery charge. We now focus on the dynamics of the charging process, assuming that both battery and charger are initialised in their respective ground states.  First we consider the case where the control is set to optimally stabilise the charger's population inversion, $f=\Gamma$. Some representative results for the time-dependent energy and ergotropy are shown in Fig.~\ref{fig:charge_dynamics}, obtained by numerically solving the master equation~\eqref{unconditional_master_equation} with efficiency $\eta = 0.3$.  We see that both the energy and ergotropy grow monotonically towards their asymptotic value. This highlights the stability of the charging process, meaning that the precise instant at which the battery is extracted is unimportant so long as sufficient time has elapsed.

To emphasise the random character of the underlying measurement and feedback process, we include in Fig.~\ref{fig:charge_dynamics} some trajectories obtained by solving the stochastic master equation~\eqref{drive_after_measure} for the same parameters. We numerically integrate Eq.~\eqref{drive_after_measure} using the Euler-Milstein scheme proposed in Ref.~\cite{Rouchon2015}, which guarantees complete positivity. Each trajectory represents a possible outcome of a single run of the charging protocol. The battery energies for different trajectories show a significant dispersion during the transient energy transfer process, yet these fluctuations are strongly suppressed in the steady state by the stabilising effect of the feedback. This indicates that the charger works not only in an ensemble-averaged sense but also at the single-trajectory level, even at low measurement efficiency. This remarkable effectiveness is partly due to the assumption of Markovian feedback: the performance will be seen in Sec.~\ref{sec:delay} to deteriorate substantially when large time delays in the feedback loop are considered.

\begin{figure}
	\includegraphics[width=0.95\linewidth]{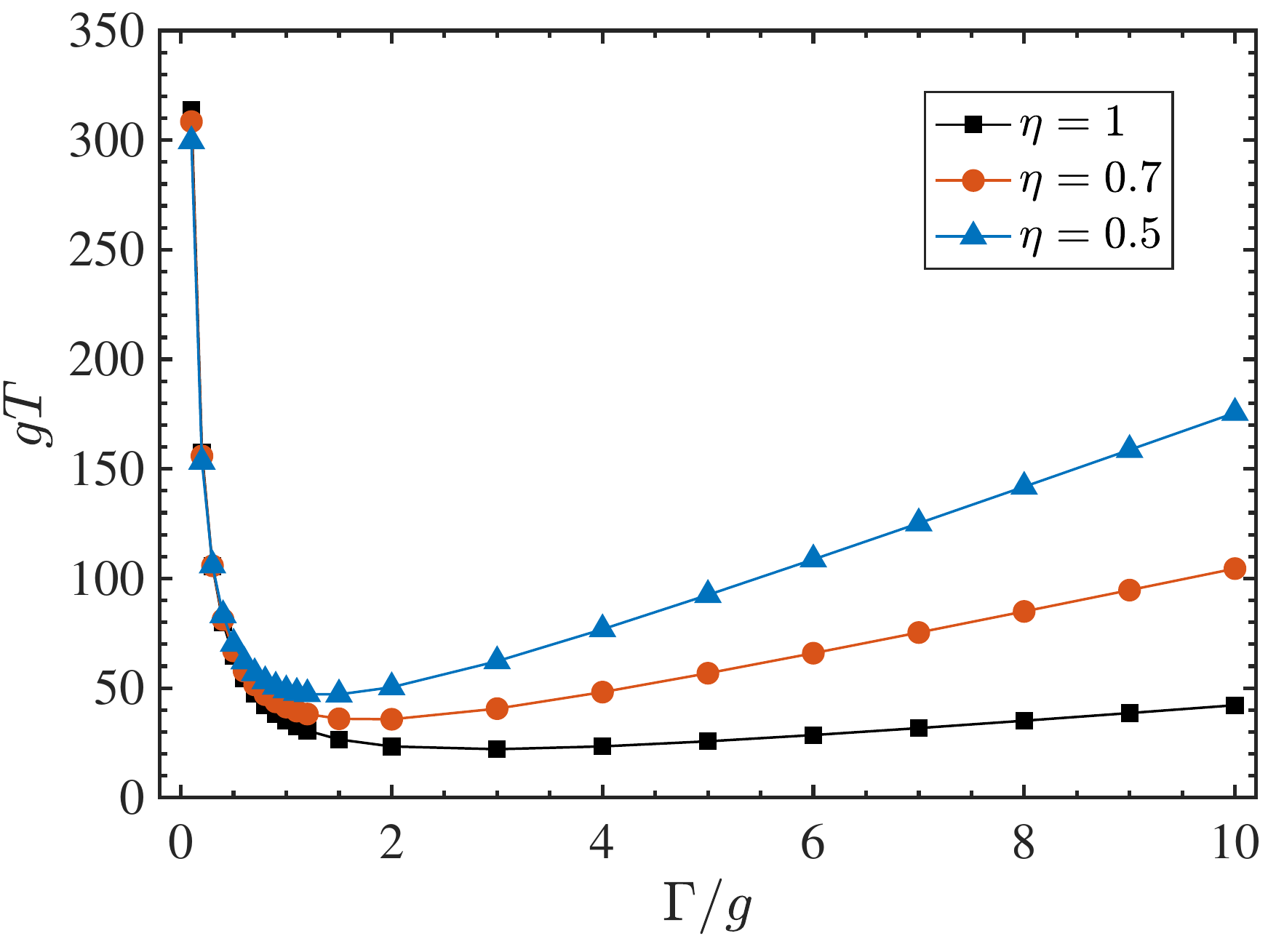}
	\caption{Total charging time $T$ for optimal Markovian feedback as a function of the dissipation and drive strength $f=\Gamma$, with $\Omega_0=0$ and $d=20$. The charging time defined by Eq.~\eqref{charging_time} is plotted as a function of the driving and dissipation strength, $f = \Gamma$, for three different values of the measurement efficiency, $\eta$. Symbols show numerical data joined up by lines to guide the eye. \label{fig:time_eta}}	
\end{figure}

For a fixed efficiency, the time taken for the battery to reach its maximum charge depends on a competition between the rate of dissipation $\Gamma$ and the coupling strength $g$. In order to quantify the charging time precisely, we identify the time $T$ at which the battery's energy differs from its asymptotic value by a fractional error $\epsilon$, i.e.
\begin{equation}\label{charging_time}
E[\hat{\rho}_B(T)] = (1-\epsilon)E[\hat{\rho}_B(\infty)],
\end{equation}
where we choose a small (arbitrary) value $\epsilon = 10^{-2}$. The behaviour of the charging time is shown in Fig.~\ref{fig:time_eta} as a function of the driving and dissipation rate $f=\Gamma$, for three different values of the measurement efficiency. Naturally, the charging process is fastest for efficient measurements, becoming progressively slower as $\eta$ is reduced below unity. We also find that for each $\eta$ there exists an optimal value of the coupling $\Gamma$ that minimises the charging time. Such an optimum is expected, since for $\Gamma = f \ll g$ the charging speed is limited by the small power input, while for $\Gamma \gg g$ the system enters a quantum Zeno regime where energy transfer from contact to battery is inhibited by frequent spontaneous emissions. Since efficient measurements use every emitted photon to improve the feedback, the decohering effect of spontaneous emission is most dramatic at low efficiencies. The optimum driving and dissipation strength is therefore $f = \Gamma \sim g$, with the optimal $\Gamma$ decreasing with $\eta$, as shown in Fig.~\ref{fig:time_eta}.

\begin{figure}
	\includegraphics[width=0.9\linewidth]{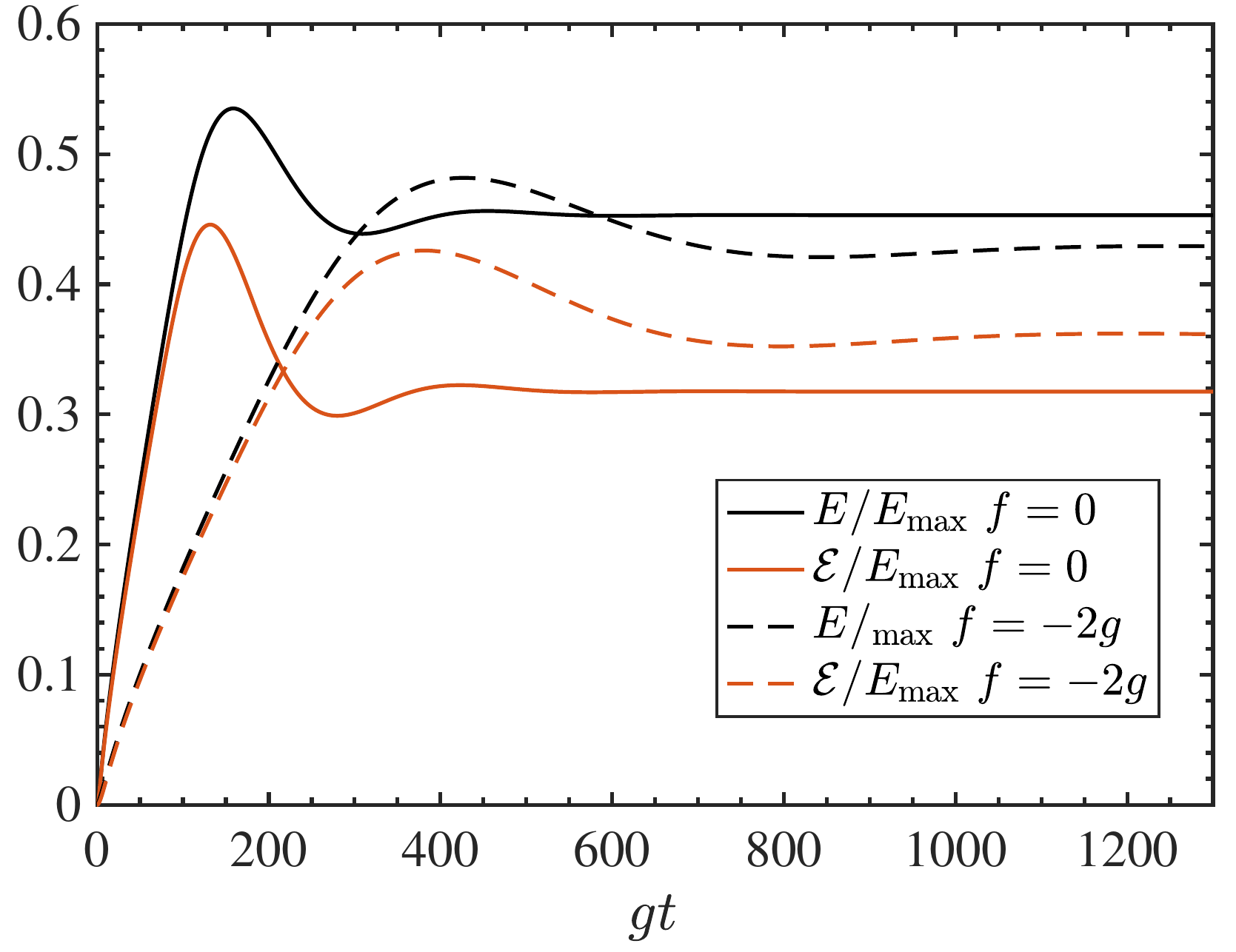}
	\caption{Ensemble-averaged charging dynamics when stabilising the charger coherence. The parameters are chosen near the optimum shown in Fig.~\ref{fig:final_charge_coherent}, with $\Gamma=10g$, $\Omega = 1.2g$, $\eta=0.3$ and $d=20$. Feedback (dashed lines) increases both the final ergotropy and the time taken to reach the stationary state, compared to a constant drive (solid lines).\label{fig:coherent_charging_dynamics}}
\end{figure}

For comparison, in Fig.~\ref{fig:coherent_charging_dynamics} we plot the battery dynamics obtained for the coherence-based charging protocol defined in Sec.~\ref{sec:stabilising_coherence}. We focus on parameters near the optimal point of Fig.~\ref{fig:final_charge_coherent} where the final ergotropy is maximised. In the absence of feedback, a constant drive is seen in Fig.~\ref{fig:coherent_charging_dynamics} to generate transient oscillations, with the ergotropy peaking at short times before settling to a smaller steady-state value. Feedback tends to reduce the magnitude of these oscillations and stabilise asymptotic states with greater ergotropy. However, the feedback also substantially increases the time taken to reach the stationary state. Overall, we observe that the timescale of the coherence-based charging protocol seen in Fig.~\ref{fig:coherent_charging_dynamics} is significantly longer than in Figs.~\ref{fig:charge_dynamics} and~\ref{fig:time_eta}, because the large impedance mismatch when $\Gamma \gg f,g$ suppresses energy transport.

\subsection{Effect of feedback delay}
\label{sec:delay}

So far we have assumed that the feedback control is applied instantaneously, but any real feedback loop has some delay due to the finite response time of the detector and controller. In this section, we examine how this delay influences the efficacy of battery charging, focussing on the optimal case where $f=\Gamma$. Since the dynamics is no longer Markovian, it is necessary to simulate the stochastic master equation~\eqref{drive_after_measure} explicitly~\cite{Rouchon2015} and average over many  trajectories. We find that the effect of time delay in the feedback loop is negligible so long as $\Gamma\tau\ll 1$, in accordance with previous studies~\cite{Patti2017}. As the time delay $\tau$ increases, shot-to-shot fluctuations grow and attaining numerical convergence of the trajectory average becomes increasingly demanding. Our examples are therefore restricted to relatively small time delays, $\Gamma\tau \lesssim 0.5$, in order to obtain reliable results with moderate resources. We take up to $500$ trajectories for each parameter set.

\begin{figure}
	\includegraphics[width=0.9\linewidth]{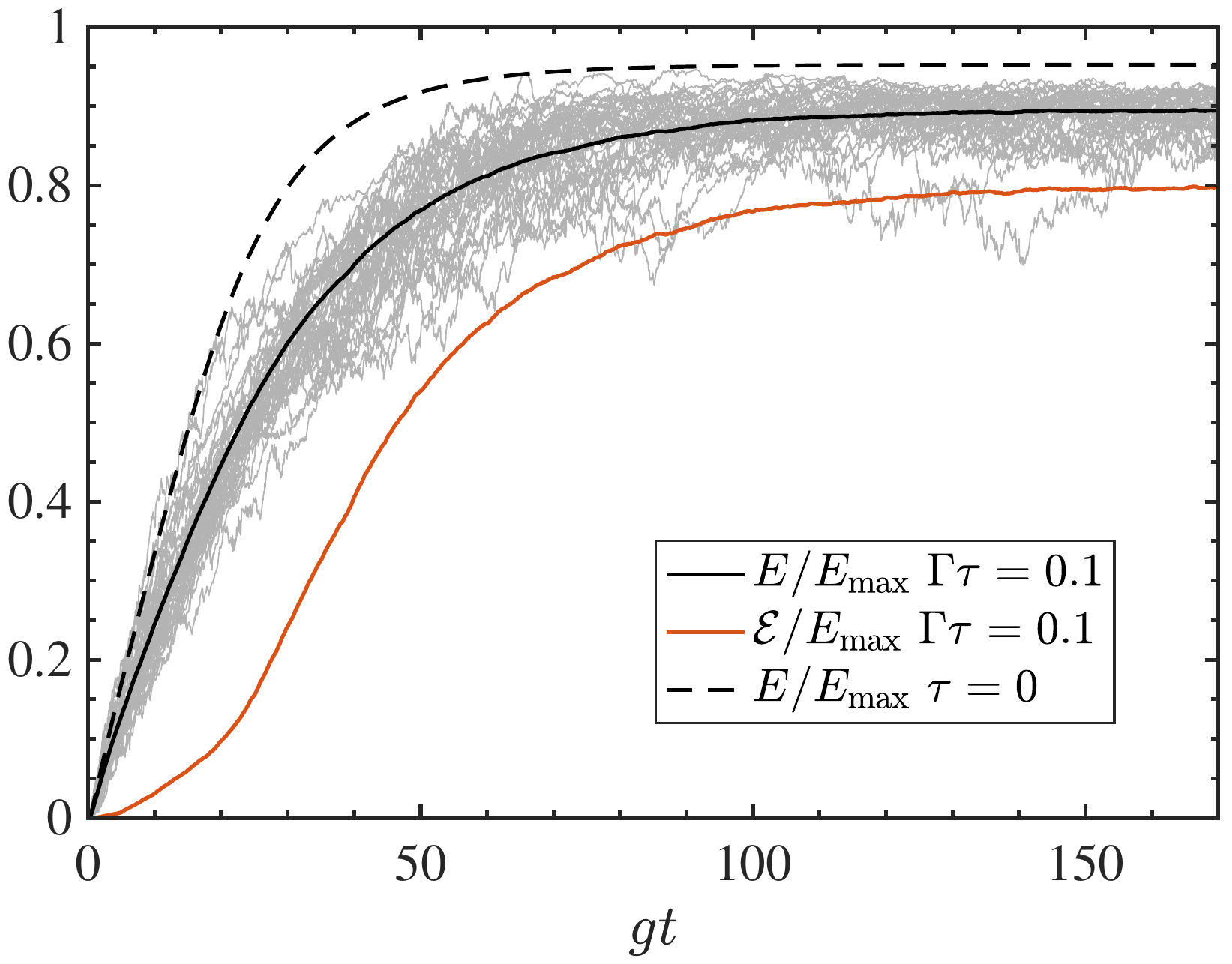}
	\caption{Charging dynamics under delayed feedback control, with $f = \Gamma = 5g$, $\Omega_0=0$, $\eta = 0.7$ and $d=10$. The ensemble-averaged battery energy (solid black line) and ergotropy (solid red line) are plotted against time together with the energy along forty individual trajectories (light grey lines) for a time delay of $\Gamma\tau=0.1$. The Markovian case (dashed black line) is shown for comparison.\label{fig:dynamics_delay} }
\end{figure}

The charging dynamics is plotted in Fig.~\ref{fig:dynamics_delay} for an example with $\Gamma\tau = 0.1$. The ensemble-averaged behaviour is qualitatively similar to the case with no delay, albeit the timescale to reach stationarity is increased as can be seen by comparing the solid and dashed black lines in Fig.~\ref{fig:dynamics_delay}. However, the battery energy along individual trajectories exhibits significant dispersion around the average even in the steady state, marking a clear departure from Markovian feedback~[c.f.~Fig.~\ref{fig:charge_dynamics}]. These fluctuations tend to increase the  entropy of the ensemble and the achievable ergotropy is correspondingly reduced. 

\begin{figure}
	\includegraphics[width=0.9\linewidth]{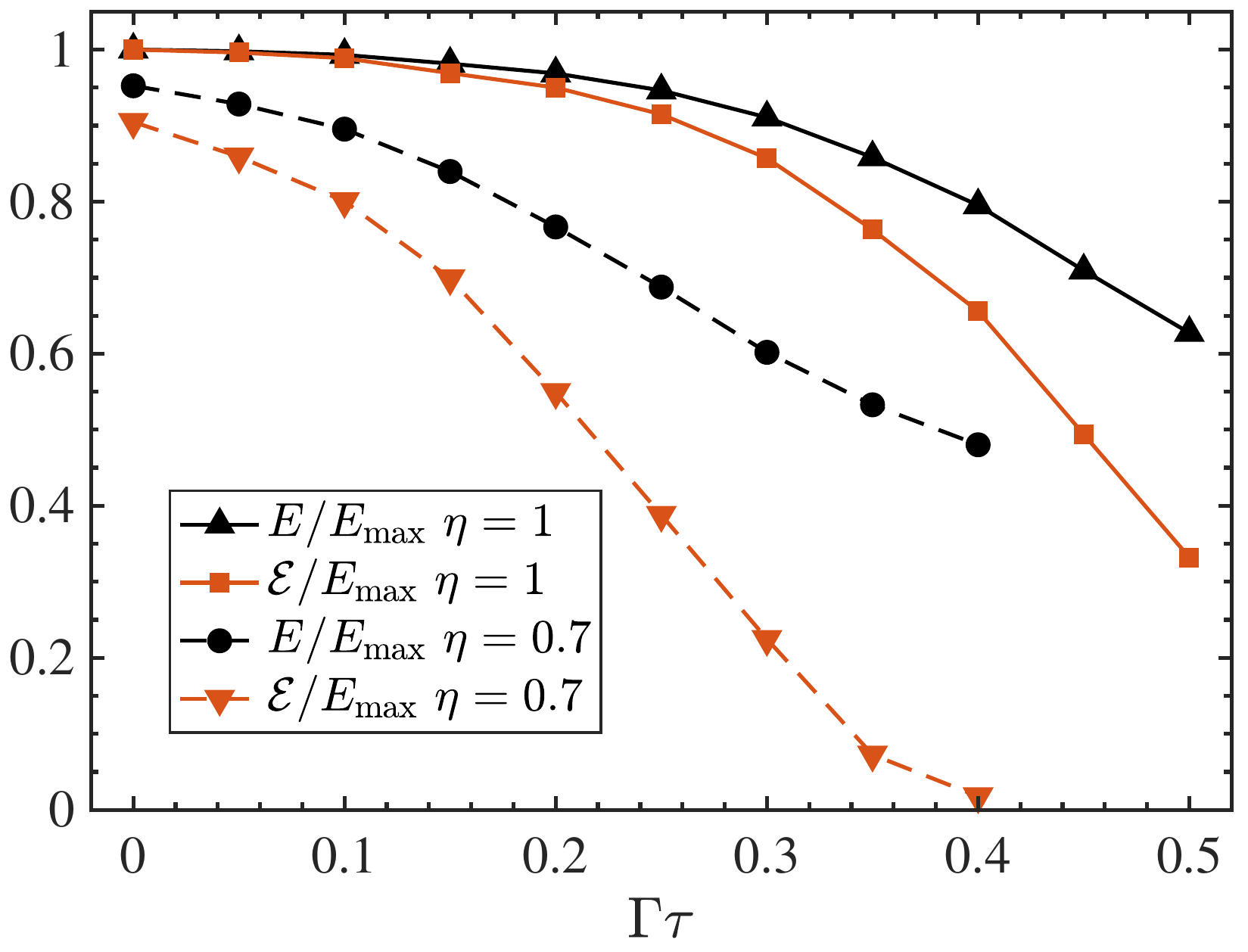}
	\caption{Steady-state battery energy (black) and ergotropy (red) as a function of the time delay, $\tau$, with $f=\Gamma=5g$, $\Omega_0=0$, and two different measurement efficiencies, $\eta$. Symbols show numerical data with conjoining lines to guide the eye.\label{fig:delay_charge}
	}
\end{figure}

To examine the effect of delay on the final battery charge in more detail, we plot the steady-state energy and ergotropy as a function of $\tau$ in Fig.~\ref{fig:delay_charge}. With efficient measurements, $\eta=1$, the delay has essentially no effect for $\Gamma\tau <0.1$, while the energy and ergotropy are seen to progressively decrease for $\Gamma\tau> 0.1$. For inefficient measurements, in contrast, the attainable charge begins to deteriorate as soon as any finite delay is introduced, as shown by the dashed lines in Fig.~\ref{fig:delay_charge}. 

As the feedback delay increases, very large differences arise between individual trajectories and the ergotropy of the ensemble-averaged state decays to zero. This indicates a complete breakdown of the feedback loop due to lag between the measurement backaction and the control response. Indeed, the measured value of $\avg{\sg_x}_r$ drifts over a timescale on the order of $\Gamma^{-1}$, by which time the delayed feedback is likely to drive the qubit away from the inverted state instead of towards it. This randomises the direction of the Bloch vector and leads the charger to a maximally mixed state, which is equivalent to infinite temperature. The battery then effectively thermalises with the qubit to a fully passive state with $E \approx 0.5E_{\rm max}$. This tendency can be seen in the dashed curves of Fig.~\ref{fig:delay_charge} for larger $\tau$.

\section{Discussion}
\label{sec:conclusion}

In this work, we have explored the use of linear feedback control to power a qubit charger coupled to a quantised, finite-dimensional battery. We have introduced two different control protocols based either on stabilising population inversion or quantum coherence, which respectively generate incoherent or coherent ergotropy. Both kinds of feedback have been shown to improve the stability and the asymptotic battery charge, as compared to the case with unconditional driving (no feedback). However, the approach of Sec.~\ref{sec:pop_inversions} based on population inversion has superior performance overall. In particular, this strategy allows for effective charging even under the realistic constraints of inefficient measurements and a small time delay in the feedback loop. Incoherent ergotropy is arguably preferable for energy storage because it is robust against dephasing and can be extracted without coherence-changing operations. In contrast, extracting work from coherent ergotropy, as generated by the protocol of Sec.~\ref{sec:stabilising_coherence}, requires a coherent drive with a definite phase relationship to the original charging field. 

Although we have simplified our model by taking the charger-battery interaction Hamiltonian to have energy-independent matrix elements, we expect our conclusions to apply in other settings, e.g.~spin systems or mechanical oscillators, whenever near-resonant interactions are a good approximation. As discussed in the Appendix, detuning between the charger and battery frequencies increases the charging time but does not affect the final charge achievable under optimal feedback. We also show in the Appendix that our scheme tolerates a small amount of finite-temperature noise. More generally, our results highlight the potential of feedback --- already well established in the context of refrigeration~\cite{Bushev2006,Tebbenjohanns2019,Conangla2019} --- for the manipulation of energy and ergotropy at the quantum scale. We have demonstrated that these quantities can be stabilised by controlling an auxiliary qubit with intuitive feedback strategies, while avoiding the need for complex time-dependent control over the battery system itself. While we have considered the simplest case of direct, linear feedback, the energetic and ergotropic capabilities of more sophisticated feedback protocols based on quantum state estimation~\cite{Doherty1999,Wiseman2002} deserve further investigation.

It is worth emphasising that the metrics we use to assess performance refer to an ensemble of many identical, independent batteries. In particular, the ergotropy only quantifies the energy that can be extracted on average. We leave the important issue of extractable work fluctuations and charging precision to future research~\cite{Friis2018,GarciaPintos2020,Crescente2020,RignonBret2021}. We also note that an experimenter could conceivably exploit their knowledge of the measurement record to optimise the work extraction step differently for each individual battery. The tools of single-shot statistical mechanics appear well suited to analyse this more involved scenario~\cite{Aberg2013,Egloff2015}.

Our main focus has been the dynamics and energetics of the charging process, leaving open several interesting questions regarding the thermodynamic value of quantum measurements~\cite{Abdelkhalek2016,Kammerlander2016,Elouard2018,Debarba2019,Guryanova2020} and feedback~\cite{Schaller2011,Vidrighin2016, Cottet2017} in this setting. Energetic constraints on discrete quantum feedback operations can be rigorously formulated by generalising the notion of entropy production to incorporate information gained by the controller~\cite{Sagawa2008,Jacobs2009,Sagawa2010,Jacobs2012,Gong2016}. Similar ideas have been applied to jump-like unravellings of open quantum system dynamics~\cite{Esposito2012,Strasberg2013a,Strasberg2013,Strasberg2017}. However, a comprehensive formulation of information thermodynamics along continuous, stochastic trajectories between quantum superposition states appears to be considerably more elusive, notwithstanding some notable recent progress~\cite{Horowitz2015,Alonso2016,Elouard2017,Dressel2017, Strasberg2019, Belenchia2020}. 

Entropic considerations aside, any serious thermodynamic account of a putative quantum battery should include the energy needed for work extraction~\cite{Monsel2020}, not to mention the power consumption of the classical control apparatus, which typically exceeds quantum scales by dozens of orders of magnitude. These problems naturally motivate~\cite{Mitchison2019} a fully autonomous approach to feedback control in quantum thermodynamics, e.g.~along the lines of Ref.~\cite{Lloyd2000}.
\\

\begin{acknowledgments}
Discussions with Bj\"orn Annby-Andersson, Javier Molina-Vilaplana, Patrick Potts, Peter Samuelsson, and Philipp Strasberg are gratefully acknowledged. M.T.M.\ and J.G.\ acknowledge funding from the European Research Council Starting Grant ODYSSEY (G. A. 758403). J.G.\ is supported by a SFI-Royal Society University Research Fellowship. J.P.\ is grateful for financial support from Ministerio de Ciencia, Innovaci\'on y Universidades (Spain) project PGC2018-097328-B-100 and Fundaci\'on S\'eneca (Murcia, Spain) Projects No. 19882/GERM/15. Calculations were performed on the Lonsdale cluster maintained by the Trinity Centre for High Performance Computing. This cluster was funded through grants from Science Foundation Ireland.
\end{acknowledgments}

\appendix

\section{Appendix: Variations of the model}

In this appendix, we analyse the role of the battery dimension and extend the model to include the effect of finite temperature and detuning. We focus for simplicity on the limit of Markovian feedback.

\subsection{Generalised model}

In general, we write the charger and battery Hamiltonians as
\begin{equation}\label{detuned_H0}
\H_C = \frac{\omega_0}{2}\sg_z,\qquad \H_B = \omega_B\hat{N},	
\end{equation}
where the energy scales $\omega_0$ and $\omega_B$ may be different. We assume the driving field is resonant with the qubit at frequency $\omega_0$. In a frame rotating at this frequency, the total Hamiltonian then reads
\begin{equation}
	\label{H_rotatingframe}
	\hat{H}(t) = \Delta_B \hat{N} + \H_{\rm int} + \H_{\rm drive}(t),
\end{equation}
with $\Delta_{B} = \omega_{B}-\omega_0$ the charger-battery detuning, and where $\H_{\rm int}$ and $\H_{\rm drive}(t)$ are given by Eqs.~\eqref{qubit_battery_interaction} and~\eqref{driving_Hamiltonian}, respectively.

To model the effect of thermal noise, we assume that the uncollected photons are emitted into a thermal radiation bath with inverse temperature $\beta$ corresponding to a mean occupation $\bar{n} = (\ee^{\beta\omega_0}-1)^{-1}$. For $\bar{n}\neq 0$, this opens the additional possibility of photon absorption by the qubit charger. The decay channel corresponding to photons collected by the detector is assumed to remain at effectively zero temperature. Therefore, the dissipative evolution is described by the superoperator (c.f.~Eq.~\eqref{heat_increment})
\begin{align}\label{dissipation_superop}
	\mathcal{K}[\hat{\rho}_r(t)] & = \eta_{\rm c} \Gamma \mathcal{D}[\sg_-]\hat{\rho}_r(t)\dd t + \sqrt{\eta \Gamma} \mathcal{H}[\sg_-]\hat{\rho}_r(t)\dd w(t) \notag \\
	&  \quad + (1-\eta_{\rm c})\mathcal{L}_{\rm th} \hat{\rho}_r(t)\dd t,
\end{align}
where $\mathcal{L}_{\rm th} \bullet = \Gamma  \left\lbrace (1+\bar{n}) \mathcal{D}[\sg_-]\bullet+ \bar{n}\mathcal{D}[\sg_+] \bullet\right \rbrace  $ is the standard Lindblad dissipator describing thermal emission and absorption~\cite{Wiseman2009}. We recall that $\eta_{\rm c}$ denotes the fraction of photons collected in the detection channel, leading to a total efficiency $\eta=\eta_{\rm c}\eta_{\rm d}$, where $\eta_{\rm d}$ is the detector efficiency.

Linear, Markovian feedback is described by Eq.~\eqref{feedback_drive} in the limit $\tau\to 0$. Following the procedure described in Sec.~\ref{sec:feedback_dynamics}, we obtain a Markovian master equation for the ensemble-averaged density operator, which reads 
\begin{align}
	\label{master_equation_general}
		\od{\hat{\rho}}{t} & = -\ii [ \Delta_{B}\hat{N} +  \H_{\rm int} + \Omega_0\sg_y,\hat{\rho}] \notag \\ 
		&\quad + \ii f [\sg_y, \sg_- \hat{\rho} + \hat{\rho} \sg_+] \notag \\
	&\quad  +\eta_{\rm c} \Gamma \DD[\sg_-] \hat{\rho}  + \frac{f^2}{\eta\Gamma} \DD[\sg_y]\hat{\rho} + (1-\eta_{\rm c})\mathcal{L}_{\rm th}\hat{\rho}.
\end{align}
This reduces to Eq.~\eqref{unconditional_master_equation} in the absence of detuning, $\omega_B = \omega_0$, and in the limit of zero temperature, $\beta\omega_0\to\infty$. 

\begin{figure}
	\includegraphics[width=0.9\linewidth]{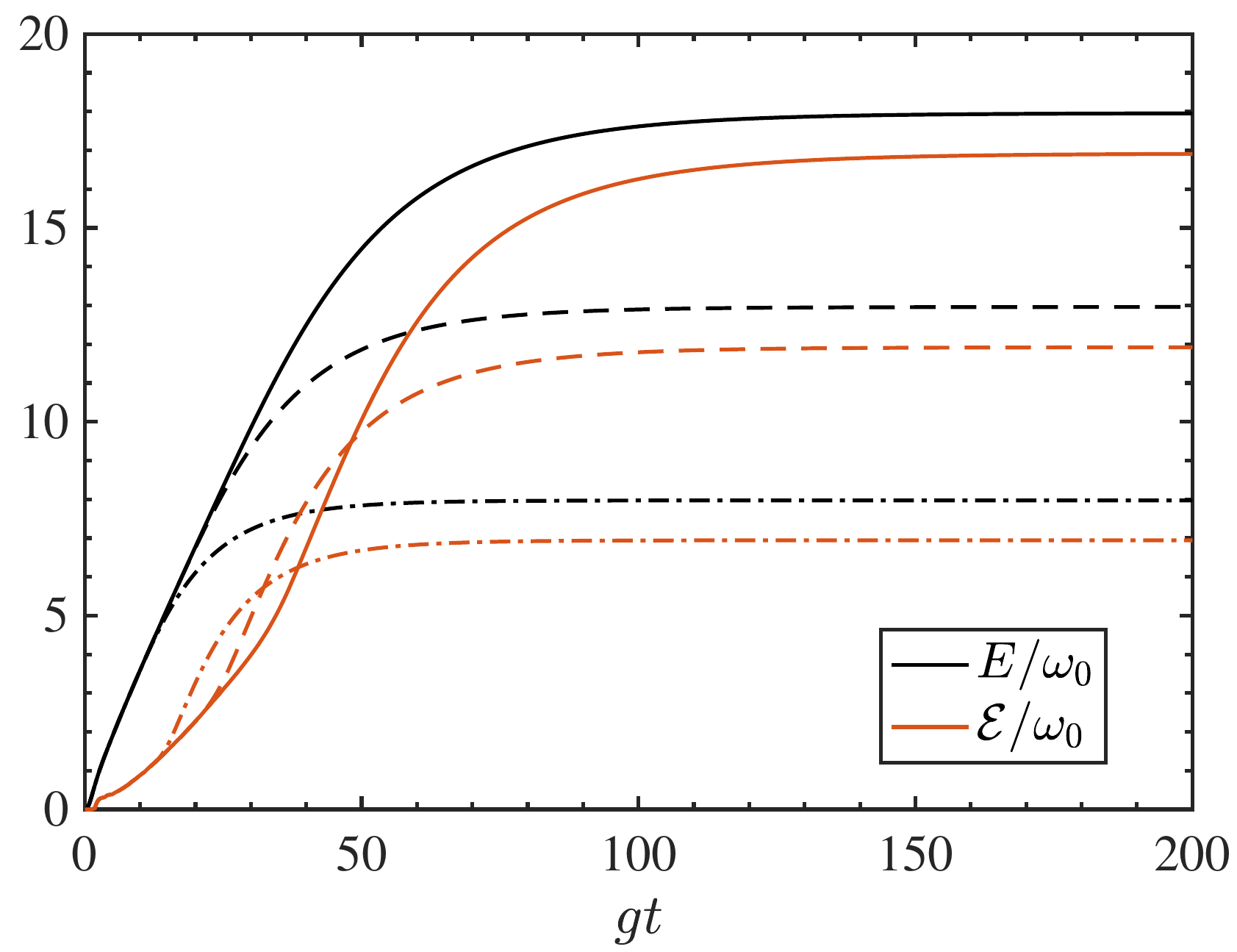}
	\caption{Evolution of the battery energy $E$ and ergotropy $\mathcal{E}$ for three different dimensions: $d=10$ (dot-dashed lines), $d=15$ (dashed lines), and $d=20$ (solid lines). We take zero temperature and parameters $\omega_B=\omega_0$, $\Gamma = f=g$, $\Omega_0=0$, and $\eta=0.5$.\label{fig:battery_dim}}
\end{figure}

\subsection{Battery dimension}

Let us first briefly address the role of the battery dimension $d$. For this analysis, it is sufficient to assume that the charger and battery frequencies are resonant, $\omega_B = \omega_0$, and take the zero-temperature limit, as in the main text. 

Fig.~\ref{fig:battery_dim} shows the time evolution of the battery charge for three values of $d$. Here, we do not normalise by $E_{\rm max} = \omega_0(d-1)$, in order to better highlight the differences between the three cases. At short times, i.e.~when $E\ll E_{\rm max}$, the battery energies and ergotropies evolve identically for all three values of $d$, while at later times the curves depart from each other due to the upper-bounded spectrum of the battery. Larger values of $d$ obviously allow for a greater steady-state energy and ergotropy, although the battery takes longer to fully charge. Overall, the evolution is qualitatively similar for all values of $d$, and this conclusion also holds for other parameter choices.

\subsection{Finite temperature and detuning}

Now we examine the effect of finite temperature, allowing also for a finite detuning, $\Delta_B\neq 0$. We focus on the control strategy where population inversion is stabilised, and thus set $\Omega_0=0$. In this case, it is possible to solve exactly for the steady state following the method described in Sec.~\ref{sec:pop_inversions}. We again posit the product ansatz $\hat{\rho} = \hat{\rho}_C\hat{\rho}_B$, with factors given formally by  Eq.~\eqref{rho_C_stationary}. It is then straightforward to check that $[\Delta_B\hat{N} + \H_{\rm int},\hat{\rho}] = 0$ for this state. The state of the charger follows from the solution of $\mathcal{L}_C\hat{\rho}_C=0$, where the dissipator $\mathcal{L}_C$ acts only on the qubit and is given explicitly by the second and third lines of Eq.~\eqref{master_equation_general}. We obtain
\begin{equation}\label{sigma_z_thermal}
	\langle \sg_z\rangle = \frac{2f-\Gamma}{\Gamma + 2\Gamma\bar{n}(1-\eta_{\rm c}) -2f +2f^2/\eta\Gamma}.
\end{equation}
The steady-state properties are manifestly independent of the detuning, $\Delta_B$, which affects only the transient dynamics under this control strategy.

\begin{figure}
	\includegraphics[width=0.9\linewidth]{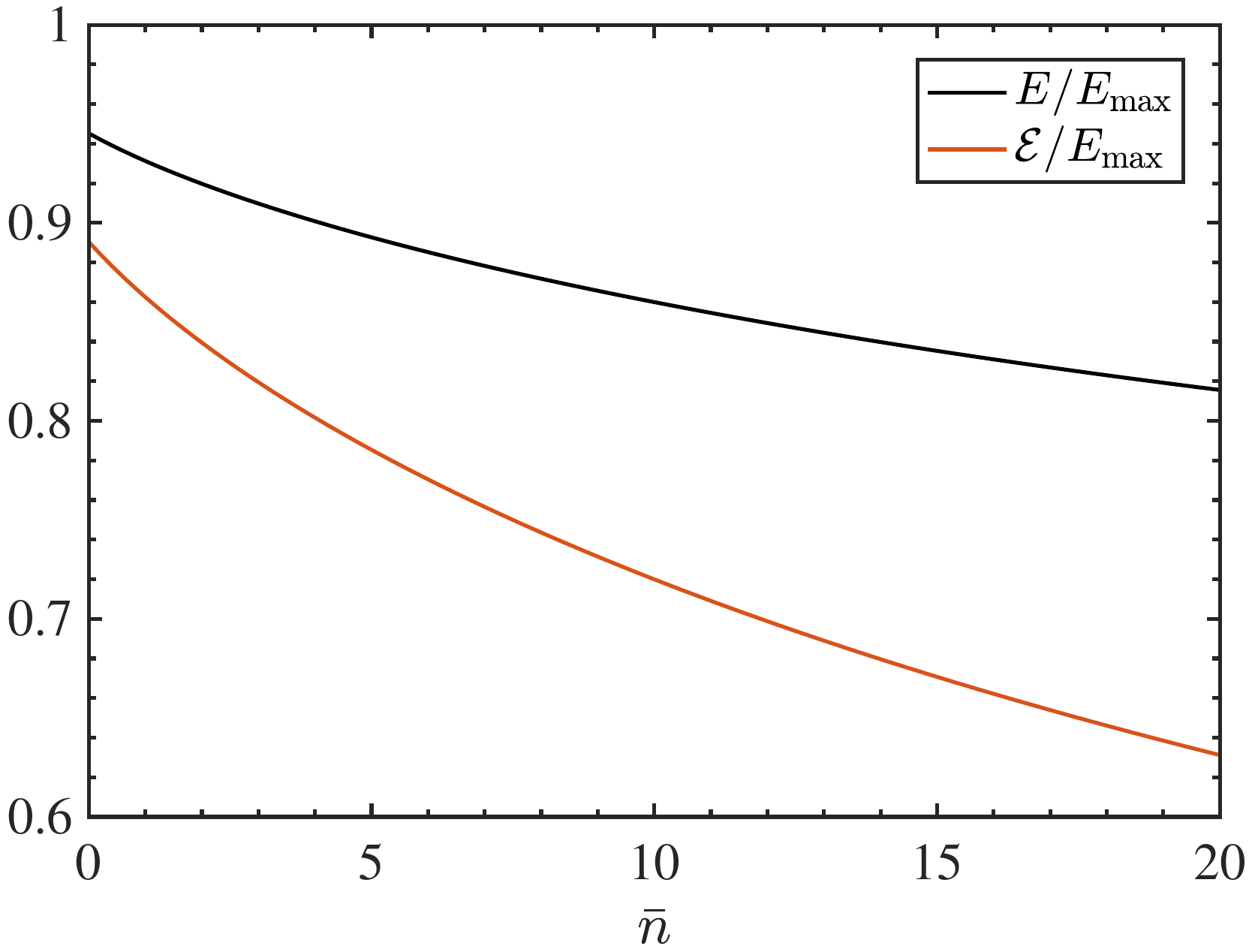}
	\caption{Steady-state battery energy $E$ and ergotropy $\mathcal{E}$ as a function of temperature, which is parametrised by the thermal occupation number $\bar{n} = (\ee^{\beta\omega_0}-1)^{-1}$, with parameters $\omega_B=\omega_0$, $\Omega_0=0$, $\eta_{\rm c}=\eta_{\rm d} = 0.7$, and $d=20$, while $f$ is tuned to the optimal value~\eqref{optimal_f_temp}. \label{fig:charge_temp} }
\end{figure}

For a given temperature and collection efficiency, the optimum feedback strength is found by maximising Eq.~\eqref{sigma_z_thermal} with respect to the ratio $f/\Gamma$. The solution is given by 
\begin{equation}\label{optimal_f_temp}
	f= \frac{\Gamma}{2} \left [1+\sqrt{1+4\bar{n}(1-\eta_{\rm c})\eta} \right ],
\end{equation}
with the corresponding population inversion (c.f.~Eq.~\eqref{sigma_max})
\begin{equation}\label{sigma_z_max_temp}
	\langle \sg_z\rangle_{\rm max} = \frac{\displaystyle\sqrt{1+4\bar{n}(1-\eta_{\rm c})\eta} + \eta - 1}{2 + 4\bar{n}(1-\eta_{\rm c}) - \eta}.
\end{equation}
Therefore, finite-temperature dissipation generally reduces the maximal population inversion of the qubit charger, and increases the feedback strength necessary to stabilise it. In particular, perfect charging is no longer possible because $\langle \sg_z \rangle <1$. The corresponding steady-state battery charge is plotted in Fig.~\ref{fig:charge_temp} as a function of the thermal occupation. The ergotropy and energy are both seen to decrease monotonically with increasing temperature. A similar reduction in charging performance with temperature is observed for other parameters, including coherent stabilisation strategies with $\Omega_0\neq 0$.

\begin{figure}
	\includegraphics[width=0.9\linewidth]{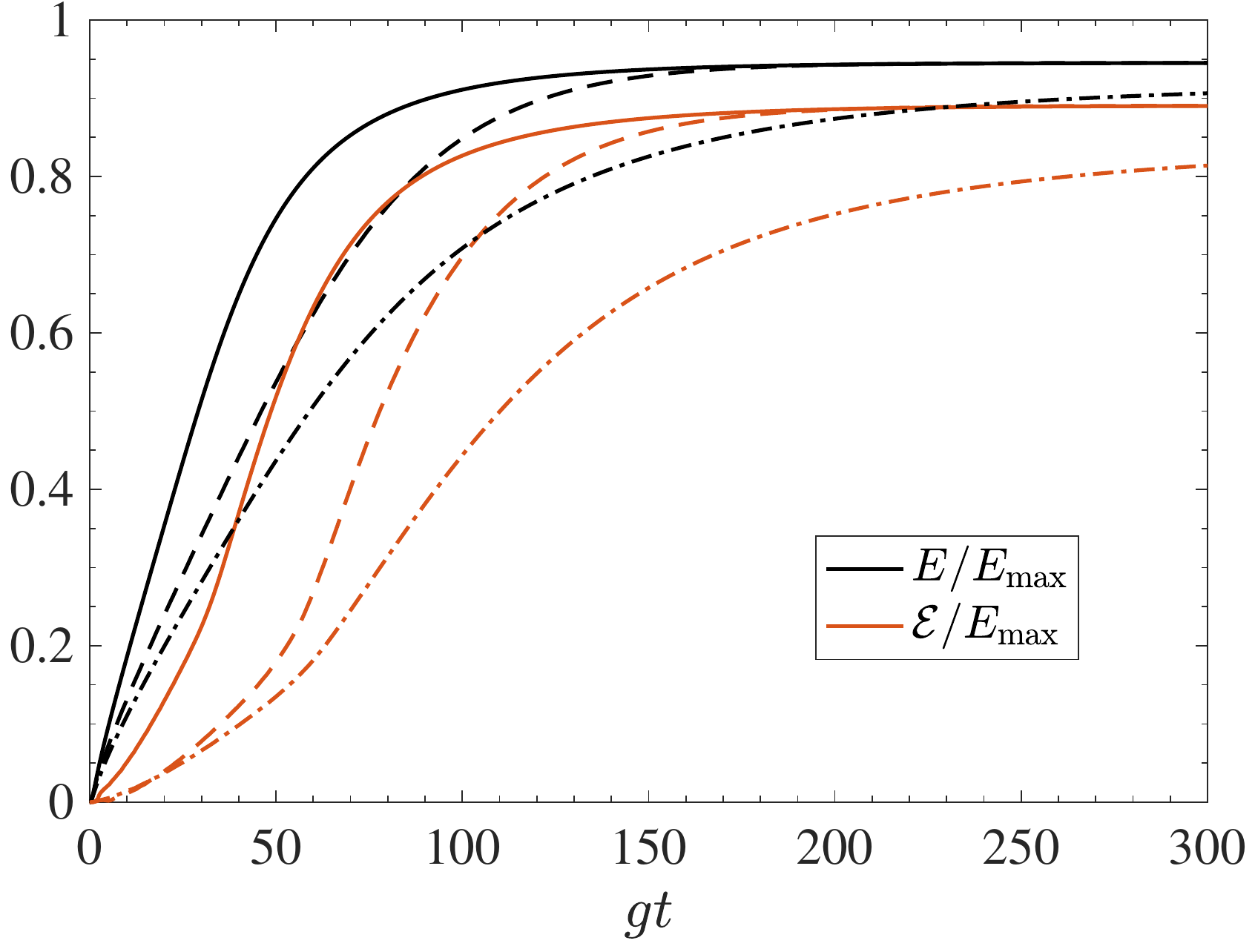}
	\caption{Dynamics of the charging process under optimal feedback, with $\Delta_B=0$ and $\bar{n}=0$ (solid lines), $\Delta_B=g$ and $\bar{n}=0$ (dashed lines), and $\Delta_B=0$ and $\bar{n}=2$ (dot-dashed lines). We also take $\Gamma=2g$, $\Omega_0 =0$ and $f$ given by Eq.~\eqref{optimal_f_temp}. 
		\label{fig:dynamics_temp_detuning}}
\end{figure}

Physically, these results can be understood in terms of the randomising effect of thermal noise. Even though thermal absorption injects energy into the charger, it does so incoherently and therefore reduces the attainable purity of the qubit state. This ultimately obstructs the coherent feedback loop from stabilising ergotropy in the battery. As a consequence, finite temperature also increases the time taken to reach the steady state. 

We illustrate the charging dynamics in Fig.~\ref{fig:dynamics_temp_detuning}, which compares the evolution of the energy and ergotropy under the optimal feedback~\eqref{optimal_f_temp} with and without thermal noise and detuning.  We see that both of these effects slow down the charging process. However, detuning has no effect whatsoever on the steady-state battery charge. Note that this conclusion holds only under the assumption of an excitation-preserving coupling in the form of Eq.~\eqref{qubit_battery_interaction}. A large value of the detuning, $\Delta_B\gtrsim g$, would activate any counter-rotating terms that are neglected when assuming an interaction of this form. These contributions open up additional transition pathways that could significantly modify the dynamics for large $\Delta_B$.

\bibliographystyle{apsrev4-1}
\bibliography{paper} 

\clearpage
\end{document}